\date{September 9, 2020} 
\documentclass[screen]{acmart}
\AtBeginDocument{%
  \providecommand\BibTeX{{%
    \normalfont B\kern-0.5em{\scshape i\kern-0.25em b}\kern-0.8em\TeX}}}

\begin{document}

\title{TeamCAD - A Multimodal Interface for Remote Computer Aided Design}

\author{Demircan Tas}
\authornote{Both authors contributed equally to this research.}
\email{tasd@mit.edu}
\orcid{0000-0002-2479-5261}
\author{Dimitrios Chatzinikolis}
\authornotemark[1]
\email{dchatzin@mit.edu}
\affiliation{%
  \institution{Massachusetts Institute of Technology}
  \streetaddress{77 Massachusetts Ave}
  \city{Cambridge}
  \state{Massachusetts}
  \country{USA}
  \postcode{02139}
}

\renewcommand{\shortauthors}{Tas and Chatzinkolis}

\begin{abstract}
Remote collaboration is a common reality of spatial design processes, but tools for computer aided design were made for single users. Via TeamCAD, we introduce a user experience where online remote collaboration experience is more like working on a table. Using speech and gesture recognition based on state of the art machine learning through webcam and microphone input, TeamCAD plugs into existing software through API's, keybindings, and mouse input.

We share results from user studies conducted on graduate students from <removed for double blind review>. Our user tests were run on Blender animation software, making simultaneous use of both modalities for given tasks. We mitigated challenges in terms of robustness and latency in readily available voice recognition models. Our prototype has proven to be an intuitive interface, providing a suitable denominator for collaborators with or without previous experience in three-dimensional modeling applications.
\end{abstract}







\maketitle

\section{Introduction}

During the COVID-19 pandemic, online conferences became the predominant method of conducting collaborative work for architectural design. Capable tools exist for verbal communication, but they are limited in capacity for tasks related to visual design activity \cite{druta2021review}. Screen sharing and annotations mitigate this problem, but there are 2 important reductions that hinder design creativity; firstly, there is a dimensionality reduction, in the sense that 3d geometries are reduced to 2d images on the screens of the users, and secondly there is a reduction in modalities, in the sense that in screen sharing, the perception is mostly visual, and in rare cases auditory. Designers, and humans in general, know more than they can tell, show, draw, model, etc. Therefore a thoughtful combination of modalities and practices is required to transfer knowledge, or communicate a design \cite{polanyi1967tacit}\cite{brave1998tangible}.

Design teams with varying levels of experience collaborate for early stage design exploration using sketches and physical models which have a natural equalizing tendency, enabling creative team work\cite{doi:10.1080/21573727.2011.569929}. This special form of collaboration is one of the key factors that add to the design process its creative aspect \cite{fussell2004gestures}. Unlike expert software running on a personal computer, everyone can actively participate in sketches and physical models without prior experience or training, resulting in a "distributed" way of designing\cite{Glăveanu2014}. Given multi-modal interaction, we argue that a similar experience will be achieved in online, remote working scenarios.

\begin{center}
    \begin{figure}[H]
      \includegraphics[width=12cm]{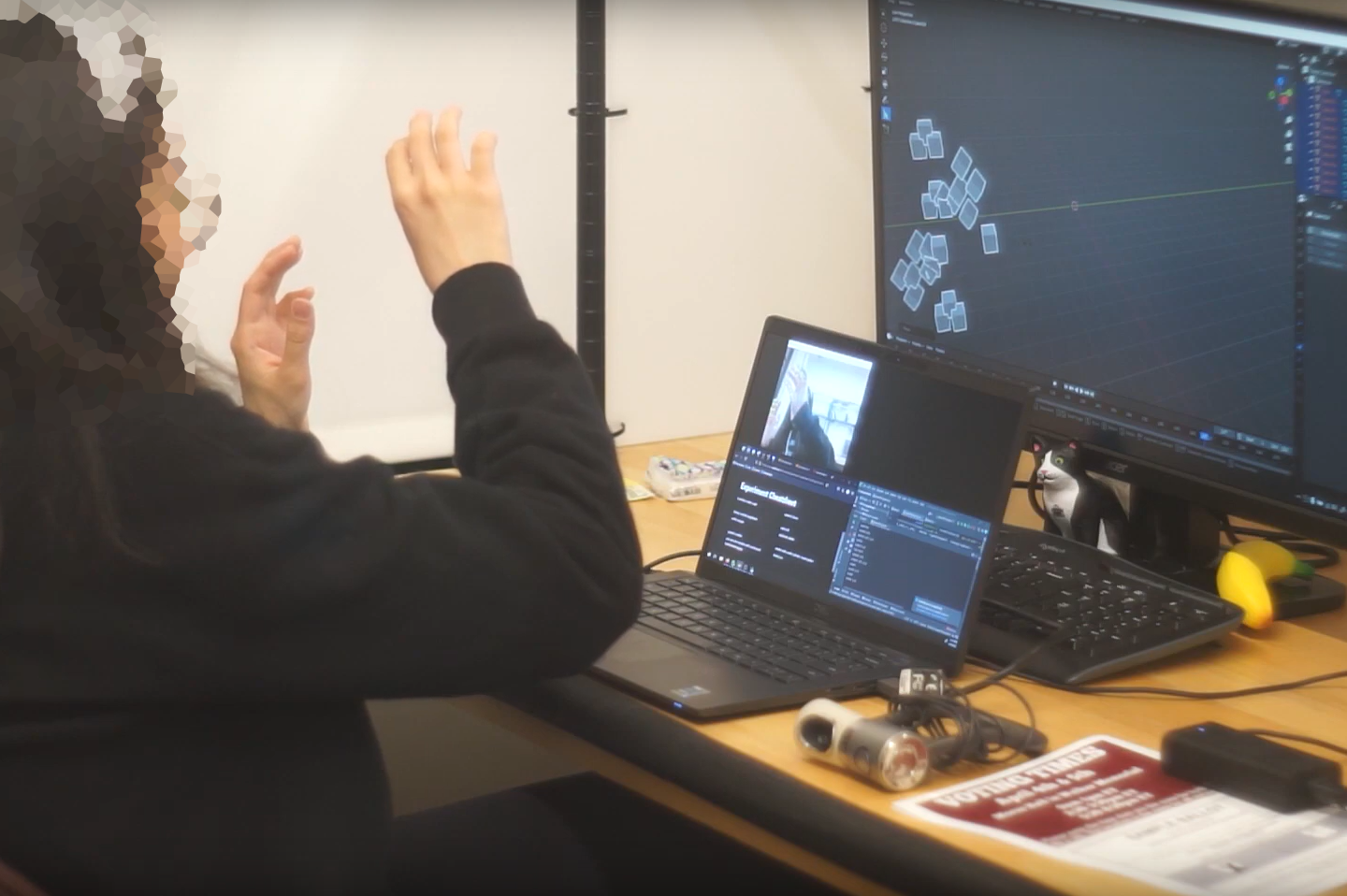}
    
      \caption{TeamCAD user testing}
      \label{fig:user}
    \end{figure}
\end{center}

We propose an interface that allows users to work on three-dimensional modeling or computer aided design (CAD) environments through speech and gesture. Our approach enables users with little prior modeling experience, while urging experienced users to elaborate their actions explicitly through voice and gestures. TeamCAD works as an interface layer, and is not limited to proprietary modeling software or plug-ins. While our implementation tests were conducted on a single computer locally, TeamCAD is capable of running remotely (subject to further tests).

\section{System Description}

TeamCAD relies on two modalities, speech and gesture. Users' voice is captured via a microphone and converted to strings via the \texttt{SpeechRecognition} Python library\footnote{\texttt{SpeechRecognition} enables access to state of the art voice to text machine learning models via their respective API's. For our system, we settled on \texttt{Google Speech Recognition} after testing CMU Sphinx, Microsoft Bing Voice Recognition, and IBM Speech to Text (\texttt{https://pypi.org/project/SpeechRecognition/}).}. The strings are queried against a dictionary of pre-defined commands using a string distance function, returning probabilities for each command. We apply probability thresholds to keep only the most likely option to infer commands from strings. We achieved our results by manually adjusting a dictionary of probability thresholds during early user testing. We also use real-time webcam images to capture commands via gestures. To this end, we use \texttt{MediaPipe}, an open-source machine learning solution developed by Google \footnote{\texttt{MediaPipe} matches skeleton structures to face, body, and hand features from imagery. These structures include three dimensional transforms for bones and joints.}. By defining preset conditions for certain events i.e. a decrease in thumb to index finger tip distance to detect \textit{pinching} we are able convert images to gesture-based commands in real-time.

Using the above mentioned techniques, TeamCAD translates voice commands and gestures to three-dimensional animation/modeling or CAD software as input. Users are provided with a heads up display (HUD) that presents the existing library of voice commands \textit{create cube, create cylinder, translate, rotate, scale}. Following a command, users' hand position dictates where the cursor, hence the selected object moves. Users can make a \textit{pinching} gesture with their right hand to select objects, and a \textit{pinching} gesture with their left hand to grab objects. Grabbed objects can be dropped by releasing the \textit{pinch} or using the \textit{enter} voice command. Ongoing transformations can be canceled using \textit{escape} command, and past actions can be undone via \textit{undo} command.

Preset translations can be triggered with commands such as \textit{greater} or \textit{smaller} as well as \textit{upwards}, and \textit{down}. Translations can also be constrained to specific axes using \textit{lateral} for $X$, \textit{lengthwise} for $Y$, and \textit{vertical} for the $Z$ axis. Throughout a transform operation, the user can utter numeric values to input precise amounts, saying ``two point one'' during \textit{vertical} scale makes an object 2.5 times as tall, saying ``forty five'' during \textit{lateral} rotation pitches the object down by 45 degrees.

\begin{center}
    \begin{figure}[H]
      \includegraphics[width=16cm]{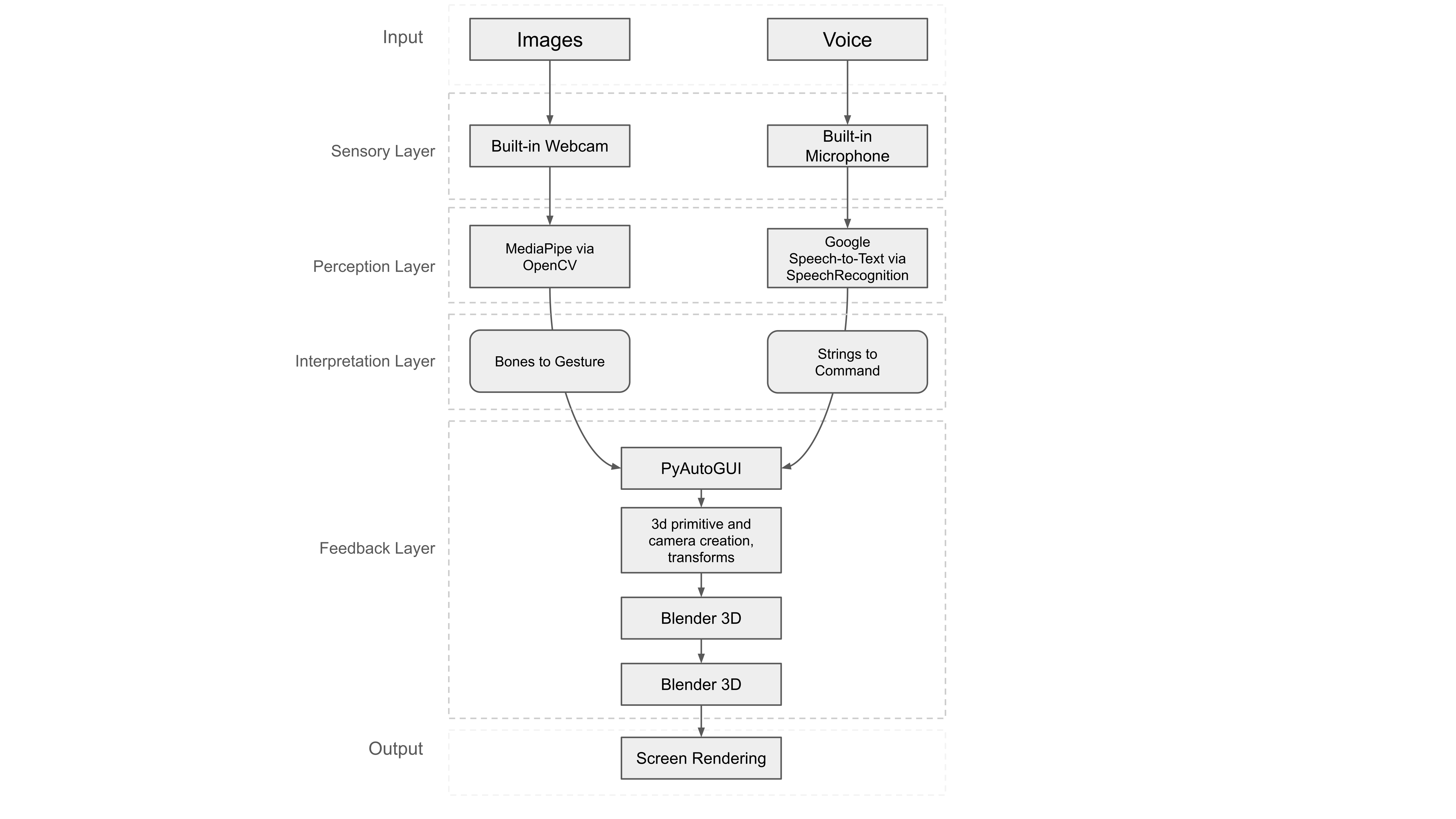}
    
      \caption{System Architecture}
      \label{fig:Architecture}
    \end{figure}
\end{center}

\section{How it Works}

TeamCAD is developed based on speech recognition and hand feature detection. 
We use the Python \texttt{SpeechRecognition} library to extract strings from users' voice. We use \texttt{MediaPipe} and \texttt{OpenCV} libraries to extract skeleton features from webcam images in real-time.

We use microphone and camera inputs for visual and audio data. Images are processed in real-time through \texttt{OpenCV} and \texttt{MediaPipe} to detect hand features (joints and connecting bones). We run \texttt{SpeechRecognition} in a parallel thread, processing voice commands in 2.5 second chunks\footnote{We achieved the best results in terms of responsiveness by limiting audio sample length to 2.5 seconds.}. \texttt{SpeechRecognition} outputs strings, while \texttt{MediaPipe} outputs lists of three-dimensional point coordinates corresponding to hand joints and finger tips.

Using \texttt{PyAutoGUI}, we translate strings, coordinates, and gesture triggers to virtual mouse and keyboard outputs and use them to manipulate objects and run commands in Blender.

\subsection{Gesture Commands}

TeamCAD relies on \texttt{MediaPipe} Python library to acquire skeleton motion data with a similar output to \cite{shotton2011real}, albeit without proprietary hardware, but from a live web camera feed instead. Instead of utilizing multiple cameras and depth input, MediaPipe works through a pre-trained neural network to infer bone transformations from a single image.

Specific configurations of bones are defined as gestures during the early stages of the project. For simplicity of use, we delegate selection and grabbing motions to separate hands. Use of bone transformations as an intermediary data structure enables users to input not only through semaphoric gestures, but also manipulative interactions\footnote{\textit{Semaphoric} in this context describes a gesture that defines a symbol i.e. a the way a person gestures to ``stop'', wherein a \textit{manipulative} gesture involves the magnitude or direction of an input i.e. pulling a virtual lever by a variable amount\cite{quek2004catchment}.}.

A limitation of this early implementation was the inability to use both hands for the tasks of selection and grabbing. After performing the initial user tests, we defined more intuitive gestures for selecting, picking, and grabbing. Both sets of index and middle fingers are mapped to a certain task, namely the right hand for moving the cursor, and the left hand for the grab command in Blender3D via \texttt{PyAutoGUI}. We mapped $X$ and $Y$ coordinates of the right hand palm landmark to the $x$ and $y$ coordinates of the mouse cursor \footnote{Capital letters refer to three dimensional coordinates, while lowercases refer to two dimensional ones}. Additionally, we defined a dynamic gesture for selecting by measuring the distance between the index fingertip and the thumb tip. When the distance is below an empirically defined threshold, we perform a ``left-click'' via \texttt{PyAutoGUI}. Similarly, by measuring the distance between the left index finger-tip and the thumb-tip, we are able to map it to the grab command of Blender3D. These minor changes in the gestural part, resulted in a more natural interaction with the computer, as the users reported in the subsequent user tests and the final demo. A section that we would like to tackle in future work is to filter the data that we are able to extract via \texttt{MediaPipe}, in order to ensure a smoother spatio-temporal coherence.

\begin{center}
    \begin{figure}[H]
      \includegraphics[width=12cm]{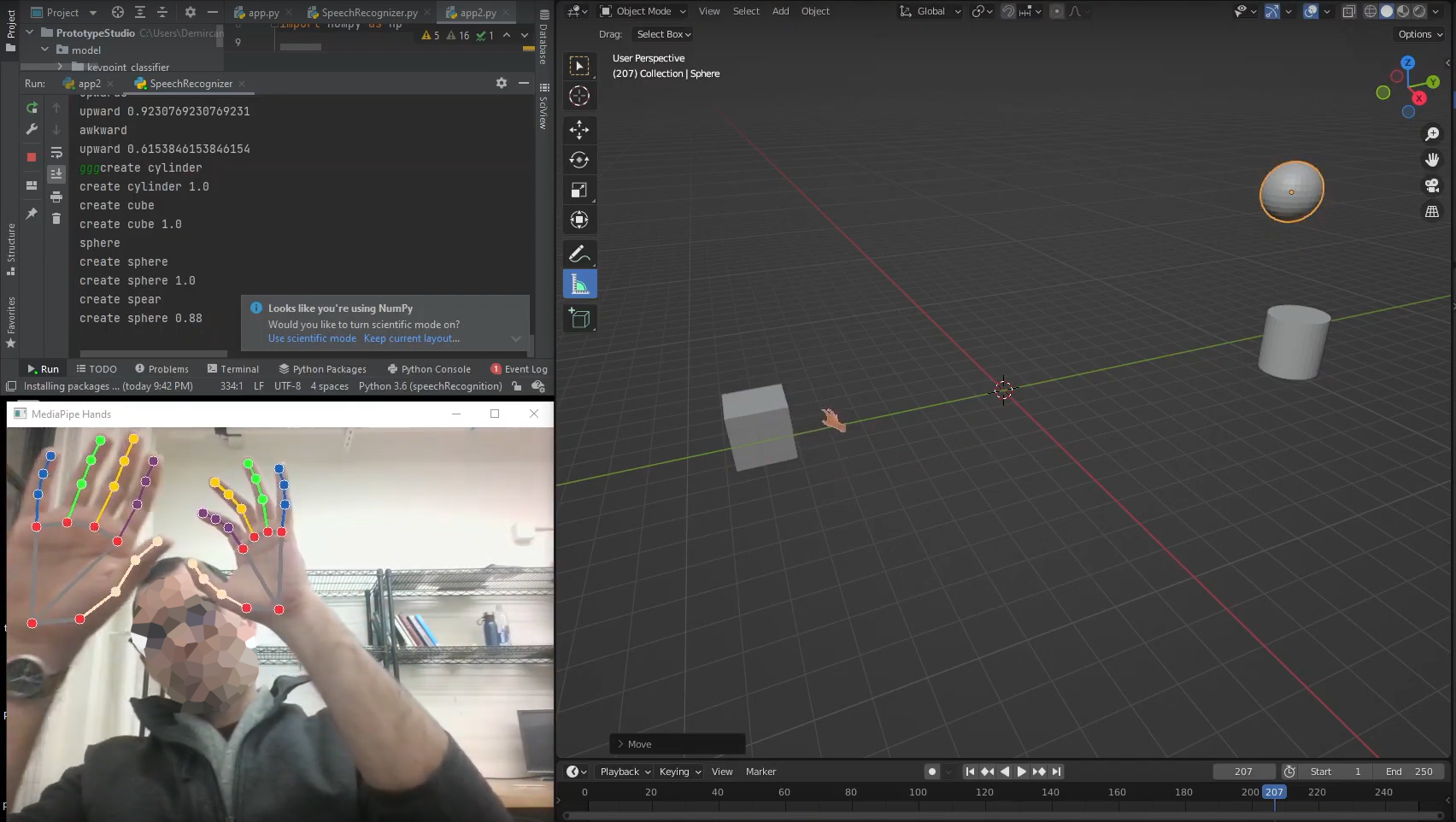}
    
      \caption{TeamCAD development demo}
      \label{fig:demo}
    \end{figure}
\end{center}

\subsection{Voice Commands}

Speech recognition works on strings returned by the \texttt{recognize\_speech\_from\_mic()} function which is wrapper for \texttt{Recognizer} and \texttt{Microphone} classes from \texttt{SpeechRecognition}. These strings are passed into the \texttt{similar()} function, which takes a target and test string, and returns a probability. We check this probability against threshold values defined for each command to decide if that command will be triggered. Thresholds for commands were adjusted manually based on user testing. We reduce a threshold if the command is poorly recognized (false negatives), and increase a threshold if the command is recognized by error (false positives).

\subsection{Pivots}

Our initial approach was to develop TeamCAD as a web based stand alone modeling interface based on Three.js. We later switched to using existing three-dimensional modeling environments in order to better focus our efforts on the implementation and improvement of modalities. By utilizing PyAutoGUI, we created a bridge among voice and gesture data, and keyboard/mouse input. This approach also provided our system with flexibility. While our studies were made using Blender, our system can be used on different software with minimal adjustment.

\section{User Studies}

We ran three iterations user studies at prototype, implementation and final studio phases. MediaPipe provided a streamlined tool, yet we implemented different methods for grabbing and selecting objects at the first two phases in order to achieve a natural feel. Moreover, we experimented with multiple setups for two handed scaling and rotating gestures. We have been unable to implement an intuitive two or three dimensional rotation without constraining axes to $X$, $Y$, $Z$ or the $camera$\footnote{This is an open problem in human computer interaction.}.

In the prototype phase we provided users with no explicit goals, other than to play around with, and get a feel for the interface. For the following implementation and final studio phases we designed an experiment where users are asked to create an arch \footnote{A curved structure of firm material, either capable of bearing weight or merely ornamental\cite{dictionary1989oxford}. The ambiguity in the spatial configurations that result in arches was intentional for our experiments, to reflect ambiguities in early stages of architectural design.} in Blender 3D software. We recorded users’ screens and camera footage as they created the requested models, and measured how much time users spend on warming up, creating objects, manipulating objects and camera movement / scene navigation. We also recorded the time users spend working through speech and gesture commands. We shuffled our pool of users for different phases of testing, and introduced new users at each new phase to avoid overfitting. Our results are from the final studio phase, where system variables have not been modified for new users.

\begin{figure}[H]
  \centering
  \includegraphics[width=0.3\linewidth]{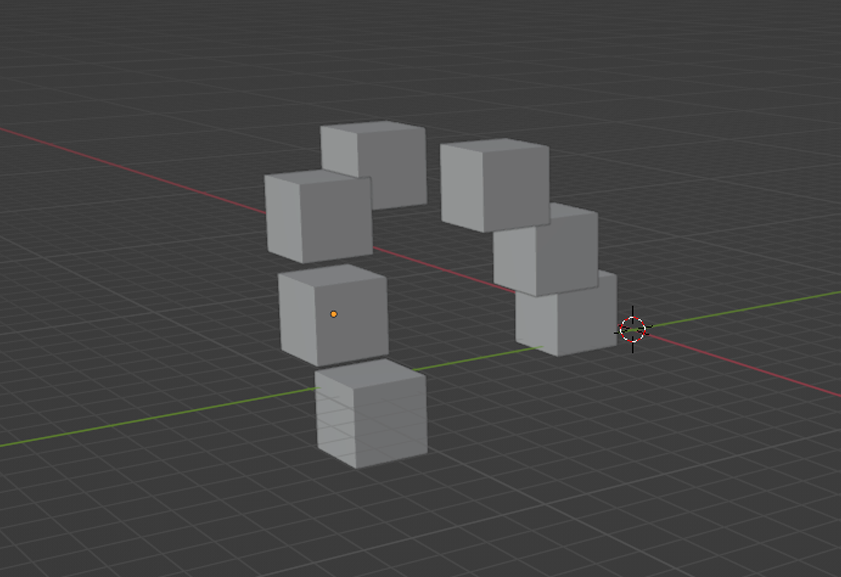}
  \includegraphics[width=0.3\linewidth]{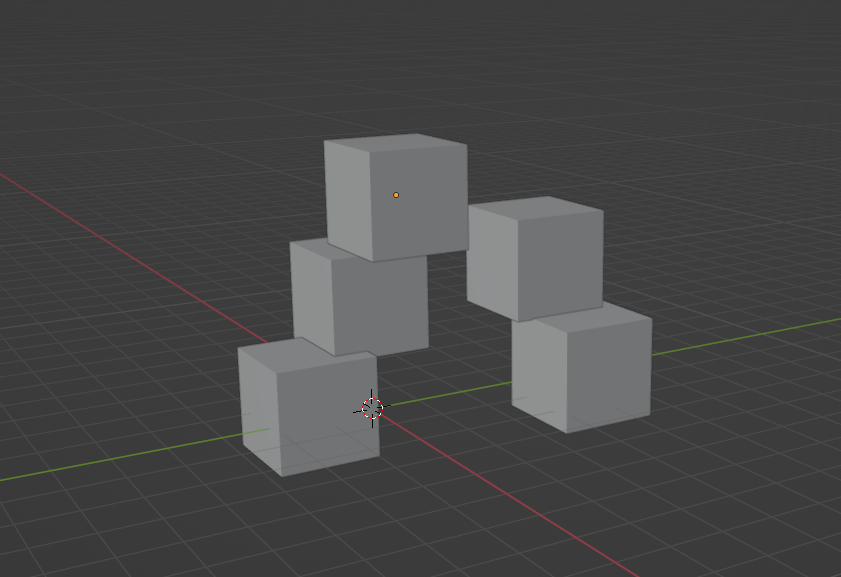}
  \includegraphics[width=0.3\linewidth]{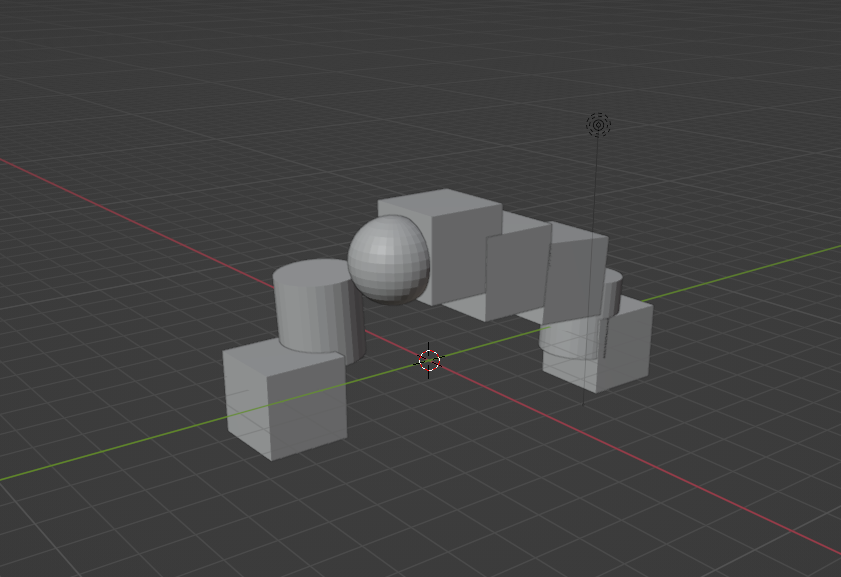}
  \includegraphics[width=0.3\linewidth]{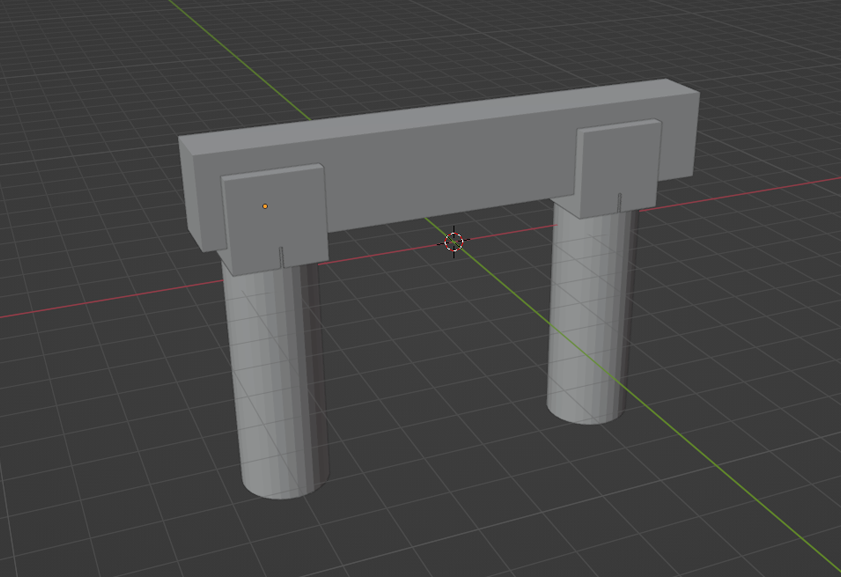}
  \includegraphics[width=0.3\linewidth]{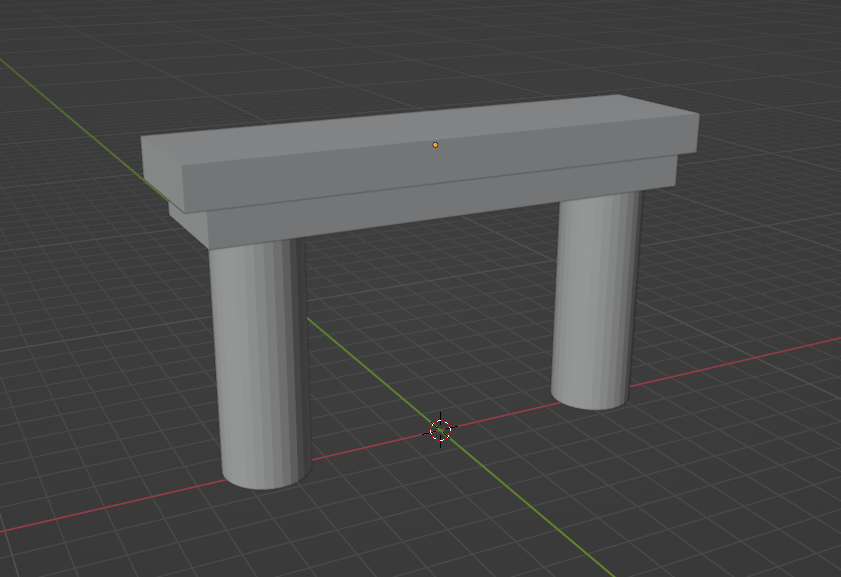}
  \includegraphics[width=0.3\linewidth]{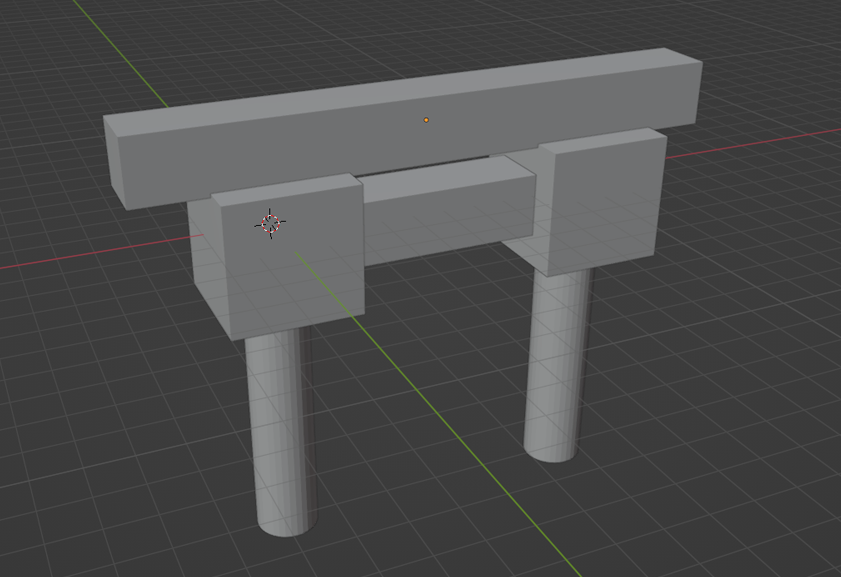}

  \caption{Arch models built by different users}
  \label{fig:arches}
\end{figure}

\section{Performance}

\begin{figure}[H]
  \centering
  \includegraphics[width=0.75\linewidth]{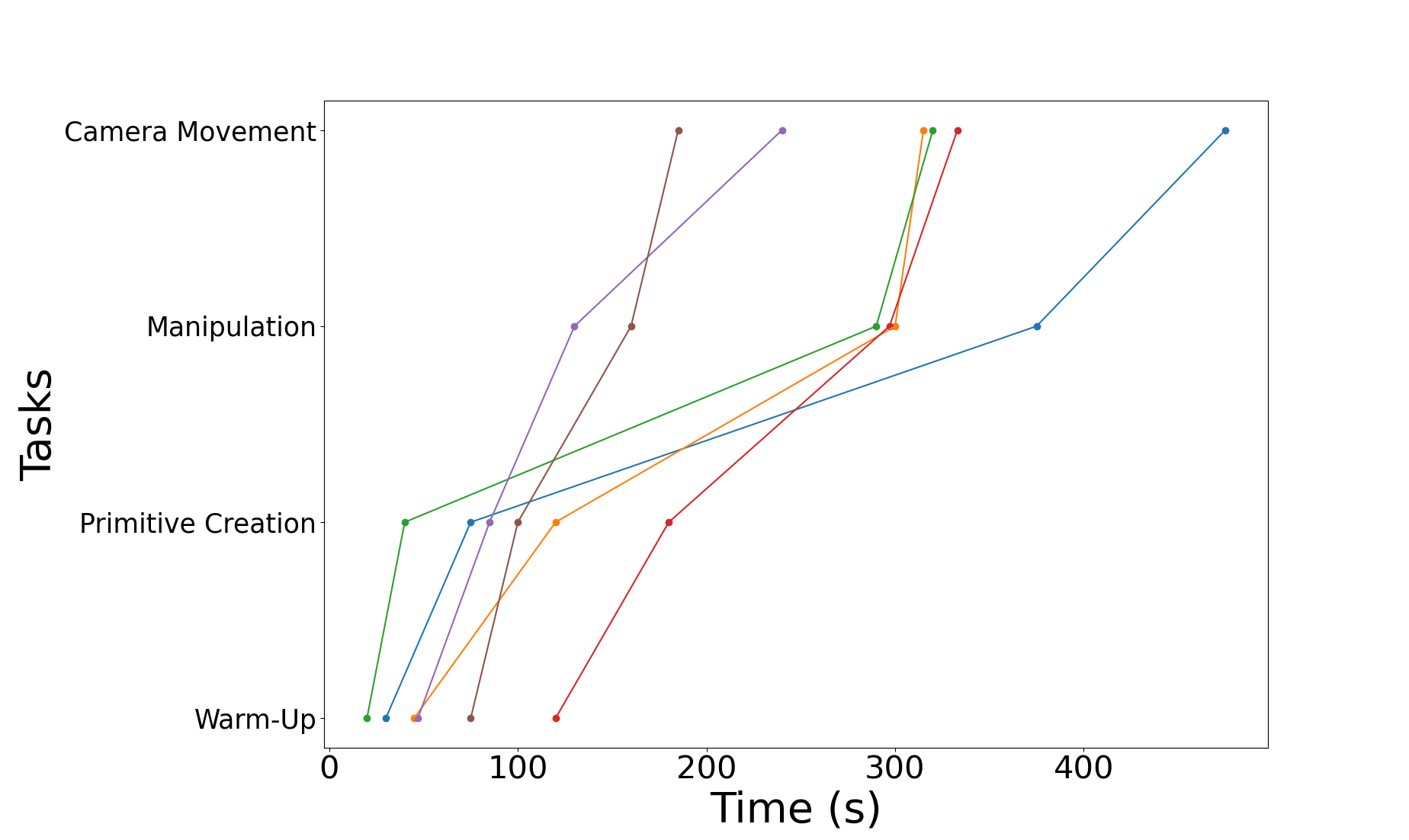}

  \caption{Time plot of different phases of modeling for each user (represented by different colors)}
  \label{fig:plot}
\end{figure}

Users spend the most amount of time on manipulating objects, while camera movement did occupy more time in cases where users got tangled in the scene. Object creation was much faster due to the library consisting of a limited number of primitives. Due to variety in previous experience, users spend different amounts of time warming up (Fig. \ref{fig:plot}).

\begin{figure}[H]
  \centering
  \includegraphics[width=0.85\linewidth]{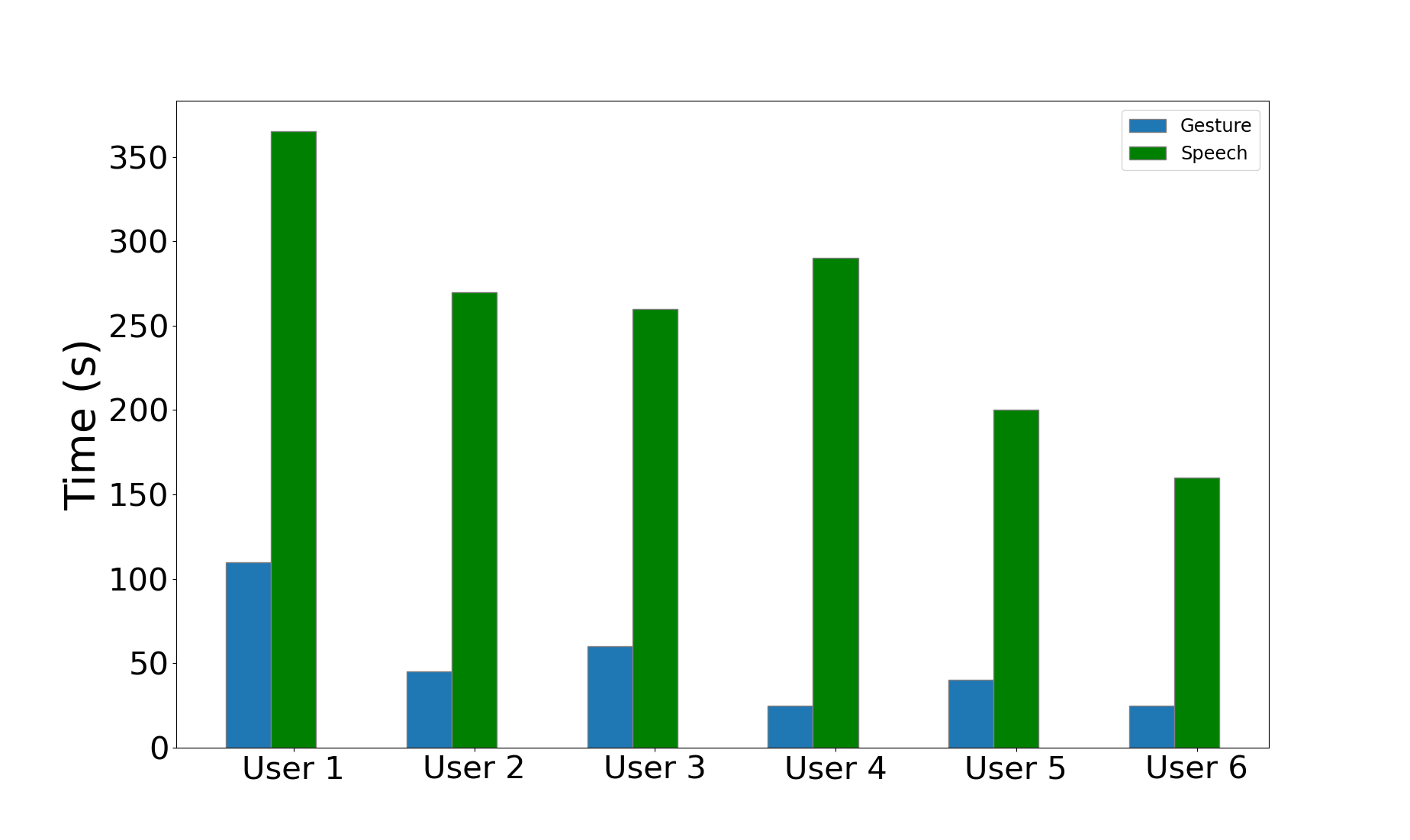}\\

  \caption{Speech versus gesture inputs based on time taken to model an arch.}
  \label{fig:hist}
\end{figure}

In terms of modality use, speech recognition occupied a substantially higher amount of time when compared to gestures. This was due to the lack of robustness and flexibility in the implemented model, Google Speech Engine. Most users struggled getting their voice commands registered, repeating a command multiple times until they found the exact pronunciation that the algorithm recognizes\footnote{Our test subjects are international students from a diverse cohort. During testing, we observed that subjects modulated their accents to trigger desired commands.} (Fig. \ref{fig:hist}). Given voice commands, we noticed additional cues of bias in the speech engine. ``front'' command was consistently recognized as ``Trump'', and sequential phrases were commonly misinterpreted as common internet search phrases. Given the performance of the speech recognition engine, we occasionally failed to strike a balance for the threshold. In such cases, we switched to words that are more distinct from other commands that trigger false positives. In some cases, this solution resulted in commands with unnatural words. Moreover, the speech engine relying on Google text-to-speech failed to provide consistent results with our experiments subjects. In most cases, users had to adjust their wording to reliably run voice. 

Almost all users had a tendency to chain commands. Since our commands were connected to distinct words, we had to specifically instruct users to leave brief gaps among commands. A partial solution was to implement longer, multiple word commands for common tasks such as \textit{create cube}, \textit{create cylinder}, etc. Blending separate commands and enabling users to chain them together is an interesting avenue of research beyond the scope of this work.

Matching our goals, users' prior experience was not consistent with how much time they spent building an arch. Using gesture with speech, users acquired a more fluid pace following the warm-up phase. The multi-modal approach had an effect of equalization in terms of time spent completing similar tasks. Some experiments even resulted in novice users completing tasks faster than those with more experience in CAD. The user that spent the least amount of time to build an arch was among the least experienced\footnote{Users were not instructed to minimize modeling time.}.

\section{Conclusion \& Discussion}

We present TeamCAD, a multi-modal interface for three-dimensional modeling and CAD. TeamCAD enables users with varying backgrounds to work intuitively on virtual modeling environments, in a similar way to physical models. We implemented TeamCAD using commonly available input hardware and state-of-the-art software for speech recognition and feature detection. While the combination of modalities provided a fluent experience for users with varying skill levels, speech recognition posed challenges that we partially overcame. While gesture recognition provided robust results with regards to different user demographics, speech recognition required extensive hand tweaking to function reliably. We solved issues of robustness by implementing a statistical approach, and manually tweaking weights based on user testing. Given more data, we can replace our hand crafted approach with machine learning models to achieve yet more robust results with less labor. Within the context of early stage architectural design modeling, the library of necessary commands were sparse enough to enable our methodology.

Based on three cycles of user studies and experiments, we have shown that TeamCAD is a common denominator for design teams consisting of members with varying skill levels. We will focus future efforts on streamlining the experience, further improving speech response, and conducting further user testing. Moreover, with access to a diverse data set, future implementations will benefit from additional stacks of multi-layer perceptrons for added performance and robustness when translating skeleton and string data to commands.


\section{Tools/Packages/Libraries}

\begin{center}
\begin{tabular}{ | p{2.85cm} | p{1.5cm}| p{2.5cm} | p{3cm} | p{3.5cm} | } 
  \hline
  Pkg/Tool/Library & Link & Purpose & Performance & Modifications \\ 
  \hline
  \small MediaPipe (0.8.10) & \tiny \href{https://mediapipe.dev/}{\path{mediapipe.dev}} & \tiny Detects skeleton features from videos / camera inputs. & \tiny It yields results comparable to Leap sensor, without the hardware and setup. Does introduce some noise. & \tiny It worked out of the box. We modified example code for
gesture recognition and better visual feedback. Our implementation strictly requires version 0.8.10. For this reason, we run gesture recognition on \textbf{Python 3.9.} \\

  \hline
  \small OpenCV (4.5.5.64) & \tiny \href{https://opencv.org/}{\path{opencv.org}}  & \tiny Provides computer vision
tools and functions in Python. & \tiny It serves as a dependency for MediaPipe to function. & \tiny Work out of the box with
no additional effort. \\

  \hline

  \small SpeechRecognition (3.8.1) & \tiny \href{https://github.com/Uberi/speech_recognition}{\path{https://github.com/Uberi/speech_recognition}}  & \tiny An easy to use speech recognition module with support for various recognition algorithms and API`s. & \tiny While SpeechRecognition worked as advertised, the algorithms and API`s that we used lacked robustness. & \tiny We had to install PyAudio as a dependency which requires Python 3.6, forcing us to run two threads using different Python versions, requiring more system resources. \\
  
  \hline

  \small PyAutoGUI (0.9.53) & \tiny \href{https://github.com/asweigart/pyautogui}{\path{github.com/asweigart/pyautogui}} & \tiny A Python library that provides mouse click and keystroke control. & \tiny Worked as intended, except for cases where hotkeys that use the ‘shift’ modifier returned capital characters instead of the combo. & \tiny Works out of the box. \\
  
  \hline
  
  \small Blender 3D (3.1.2) & \tiny \href{https://www.blender.org/}{\path{blender.org}} & \tiny Open source, enthusiast level 3d animation software & \tiny Worked as intended, except for minor issues with rotations. & \tiny We modified keyboard shortcuts to circumvent issues with ‘shift’ key modifiers. \\
  
  \hline
  
  \small PureRef (1.11.1) & \tiny \href{https://www.pureref.com/}{\path{pureref.com}} & \tiny Transparent image and text overlay for Windows 10 & \tiny Works as intended. & \tiny Works out of the box. \\
  
  \hline
\end{tabular}
\end{center}


\pagebreak

\bibliographystyle{ACM-Reference-Format}
\bibliography{sample-base}


\end{document}